\documentclass[3p,times,twocolumn]{elsarticle}
 \biboptions{comma,sort&compress}
 
\usepackage{graphicx}
\usepackage{braket}
\usepackage{amsmath,stackrel}
\usepackage{here}
\usepackage{ecrc}

\volume{00}

\firstpage{1}

\journalname{Nuclear and Particle Physics Proceedings}
\runauth{Eugenio Meg\'{\i}as et al.}

\jid{nppp}
\jnltitlelogo{Nuclear and Particle Physics Proceedings}

\usepackage{amssymb}

\usepackage[figuresright]{rotating}

\newcommand{\tr}{{\textrm {tr}}}

\newcommand{\TeV}{{\textrm {TeV}}}

\newcommand{\KK}{{\rm \mbox{{\scriptsize  KK}}}}
\newcommand{\UV}{{\rm \mbox{{\scriptsize  UV }}}}
\newcommand{\IR}{{\rm \mbox{{\scriptsize  IR }}}}
\newcommand{\SM}{{\rm \mbox{{\scriptsize  SM }}}}

\newcommand{\Imaginary}{{\textrm{Im}}}

\newcommand{\RS}{{\textrm{RS}}}

\begin{document}

\begin{frontmatter}

\title{ Continuum spectra from warped dimensions\,$^*$} 
 \cortext[cor0]{Talk presented at QCD21, 24th International Conference in QCD (5-9/07/2021, Montpellier - FR).}

\author[label1]{Eugenio Meg\'{\i}as\corref{cor1}}
\address[label1]{Departamento de F\'{\i}sica At\'omica, Molecular y Nuclear and Instituto Carlos I de F\'{\i}sica Te\'orica y Computacional, Universidad de Granada, Avenida de Fuente Nueva s/n, 18071 Granada, Spain}
\cortext[cor1]{Speaker, Corresponding author.}
\ead{emegias@ugr.es}

\author[label2]{Mariano Quir\'os}
\address[label2]{Institut de F\'{\i}sica d'Altes Energies (IFAE) and The Barcelona Institute of  Science and Technology (BIST), Campus UAB, 08193 Bellaterra, Barcelona, Spain}
\ead{quiros@ifae.es}

\pagestyle{myheadings}
\markright{ }
\begin{abstract}
\noindent
Using a five dimensional (5D) warped model with two branes along the extra dimension, we study the Green's functions for gauge bosons with a mass gap $m_g = \rho/2$ and a continuum for $s > m_g^2$. We find that the Green's functions exhibit poles in the second Riemann sheet of the complex $s$ plane, denoting the existence of broad resonances. The positivity of the spectral functions is also analyzed. Finally, it is shown how the Green's functions can modify some Standard Model processes in $pp$ collisions.
 
\begin{keyword}  Physics beyond the standard model, extra-dimensional models, Green's function, spectral function


\end{keyword}
\end{abstract}
\end{frontmatter}
\section{Introduction}

The Standard Model (SM) of particle physics has been put on solid grounds by the experiments performed at, e.g., the Large Electron Positron (LEP) or the Large Hadron Collider (LHC)~\cite{ParticleDataGroup:2020ssz}. However, it fails to describe a number of theoretical and experimental issues such as the existence of dark matter or the electroweak (EW) hierarchy problem, which demand some ultraviolet (UV) completion of the theory. In the Randall-Sundrum (RS) model~\cite{Randall:1999ee}, the hierarchy between the Planck and the EW scale is generated by a warped extra dimension in anti-de Sitter (AdS$_5$) space. Associated with each SM field, the theory predicts a discrete spectrum made out of Kaluza-Klein (KK) states with masses $m_{\KK} \sim \TeV$.

Up to now, direct searches of new physics at~colliders have been based on the detection of bumps in the in\-variant mass of final states. The elusiveness of isolated and narrow resonances led researchers to study different solutions. These include the existence of broad reso\-nances~\cite{Escribano:2021jne}, or a (quasi)-continuum of KK states, which is the idea behind the clockwork models~\cite{Giudice:2017fmj} and the~linear dilaton models (LDM)~\cite{Antoniadis:2011qw,Cox:2012ee}. A new scenario~has been recently proposed, in which there is a TeV mass gap and a continuum of states heavier than the gap. These models are characterized by the existence of an admissible metric singularity in the infrared (IR)~\cite{Csaki:2018kxb,Megias:2019vdb,Megias:2020cpw,Megias:2021arn}, and they can be considered  as modelizations of 4D theories with a gapped continuum spectrum, as can be the case of unparticles and unhiggs~theories~\cite{Georgi:2007ek,Delgado:2007dx,Falkowski:2009uy,Bellazzini:2015cgj}.

In the present work we explore the gapped continuum scenario, and provide further details on the properties of the Green's functions in the complex plane, study the  positivity of the spectral function, and analyze some phenomenological aspects in $pp$ collisions.

\section{The extra-dimensional model}
\label{sec:model}

Theories with a warped geometry are characterized by a 5D metric~$ds^2 = g_{MN}dx^M dx^N\equiv {\bar g_{\mu\nu}} dx^\mu dx^\nu-dy^2$, where $y$ is the extra dimension, and $\bar g_{\mu\nu} = e^{-2A(y)} \eta_{\mu\nu}$ is the 4D induced metric. In this section we will introduce the warped model that will be used in the rest of this work, which is analogous to the usual RS setup~\cite{Randall:1999ee}.

\subsection{General formalism}
\label{subsec:general_formalism}

We consider a scalar-gravity system with two branes at values $y=y_0$ (UV brane) and $y=y_1$ (IR brane) where we are conventionally fixing $y_0 = 0$ and $A(y_0) = 0$. The 5D action of the model reads~\cite{Cabrer:2009we}
\begin{eqnarray}
\hspace{-1.3cm} &&S = \int d^5x \sqrt{|\det g_{MN}|} \left[ -\frac{1}{2\kappa^2} R + \frac{1}{2} (\partial_M \phi)^2 - V(\phi) \right] \nonumber \\
\hspace{-1.3cm} && \qquad\qquad - \sum_{\alpha} \int_{B_\alpha} d^4x \sqrt{|\det \bar g_{\mu\nu}|} \lambda_\alpha(\phi) + S_{\!\textrm{GHY}}  \,.
\end{eqnarray}
There are three kinds of contributions: the bulk, the branes and the Gibbons-Hawking-York term.  $V(\phi)$ is the bulk scalar potential, $\lambda_\alpha(\phi)$ are the UV $(\alpha = 0)$ and IR $(\alpha = 1)$ 4D brane potentials, which fix $\phi(y_\alpha)\equiv \phi_\alpha$, and $\kappa^2 = 1/(2M_5^3)$ with $M_5$ being the 5D Planck scale.  The equations of motion (EoM) in the bulk can be written in terms of the superpotential, $W(\phi)$, as~\cite{DeWolfe:1999cp}
\begin{equation}
\phi^\prime(y) = \frac{1}{2} W^\prime(\phi) \,, \qquad A^\prime(y) = \frac{\kappa^2}{6} W(\phi) \,,
\end{equation}
while the brane potentials impose some boundary and jumping conditions in the branes. In order to solve the hierarchy problem, the brane dynamics should fix $(\phi_0,\phi_1)$ to get $A(\phi_1) - A(\phi_0) \approx 35$, and this implies $M_{\textrm{Planck}} \simeq 10^{15} M_{\textrm{TeV}}$. This formalism has been extensively discussed in the literature, see e.g. Ref.~\cite{Csaki:2000zn}.

\subsection{The gapped continuum model}
\label{subsec:softwall_model}

 A simple soft-wall model solving the hierarchy problem defined by $W(\phi) = \frac{6k}{\kappa^2} \left(1 +  e^{\nu \phi} \right)$ was introduced in Refs.~\cite{Cabrer:2009we,Csaki:2018kxb,Megias:2019vdb}. The properties of the spectrum of the KK modes of the fields depend on the value of the parameter $\nu$. It turns out that there is a critical value $\nu_c \equiv \kappa/\sqrt{3}$ such that one can distinguish between three different situations: i) $\nu < \nu_c$ corresponding to ungapped continuum KK spectra similar to unparticles; ii) $\nu = \nu_c$ corresponding to  continuum KK spectra with a gap; and iii) $\nu > \nu_c$ which leads to discrete KK spectra~\cite{Cabrer:2009we}. In the following we will focus on the critical case $\nu = \nu_c$. 

We will approximate the exact metric of Ref.~\cite{Megias:2019vdb}, by~\cite{Megias:2021arn}
\begin{equation}
\hspace{-0.6cm} A(y) =  ky \Theta(y_1 - y) + \left[ k y_1 - \log \left( k(y_s - y) \right)  \right] \Theta(y - y_1) \,  \label{eq:Ay}
\end{equation}
where $\Theta(z)$ is the step function. This simplified metric allows to get analytical expressions for the Green's functions. The geometry near the UV is AdS$_5$, while there is an admissible singularity at $y = y_s$. Using conformally flat coordinates, $dy = e^{-A(z)} dz$, the behavior in the range $z \in [z_1,+\infty)$ with $z_1 = 1/\rho$, is
\begin{equation}
\phi(z) \simeq \frac{1}{\nu_c} \rho \, z  \quad \textrm{and} \quad A(z) \simeq \rho\, z  \,, \label{eq:phi_A_z}
\end{equation} 
where $\rho = k \, e^{-A_1} \sim \TeV$ with $A_1 \equiv A(\phi_1)$. This behavior is common with the LDM~\cite{Antoniadis:2011qw,Cox:2012ee,Megias:2021mgj}.

\section{Green's functions for massless gauge bosons}
\label{sec:Green_functions}

\begin{figure*}[ht]
\centering
\includegraphics[width=4.8cm]{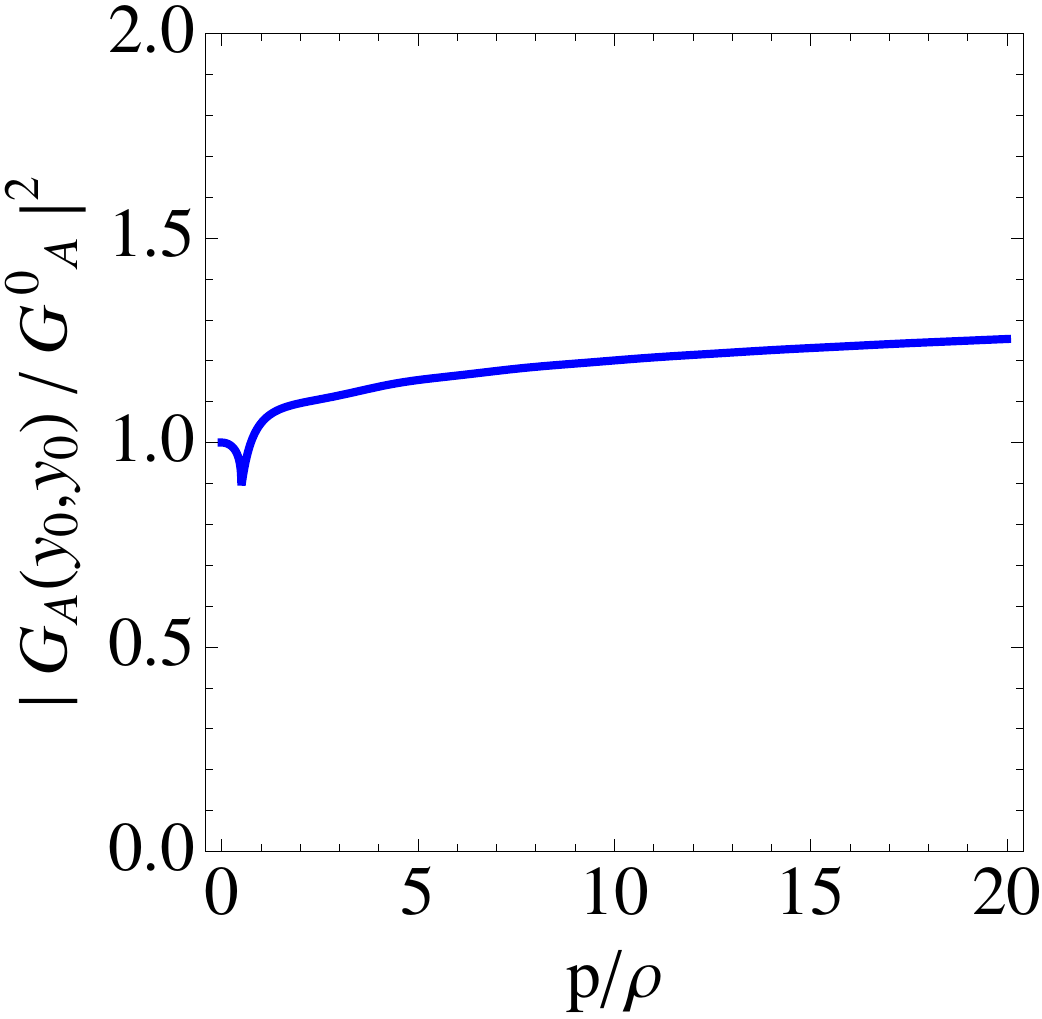} \hspace{0.5cm}
\includegraphics[width=4.6cm]{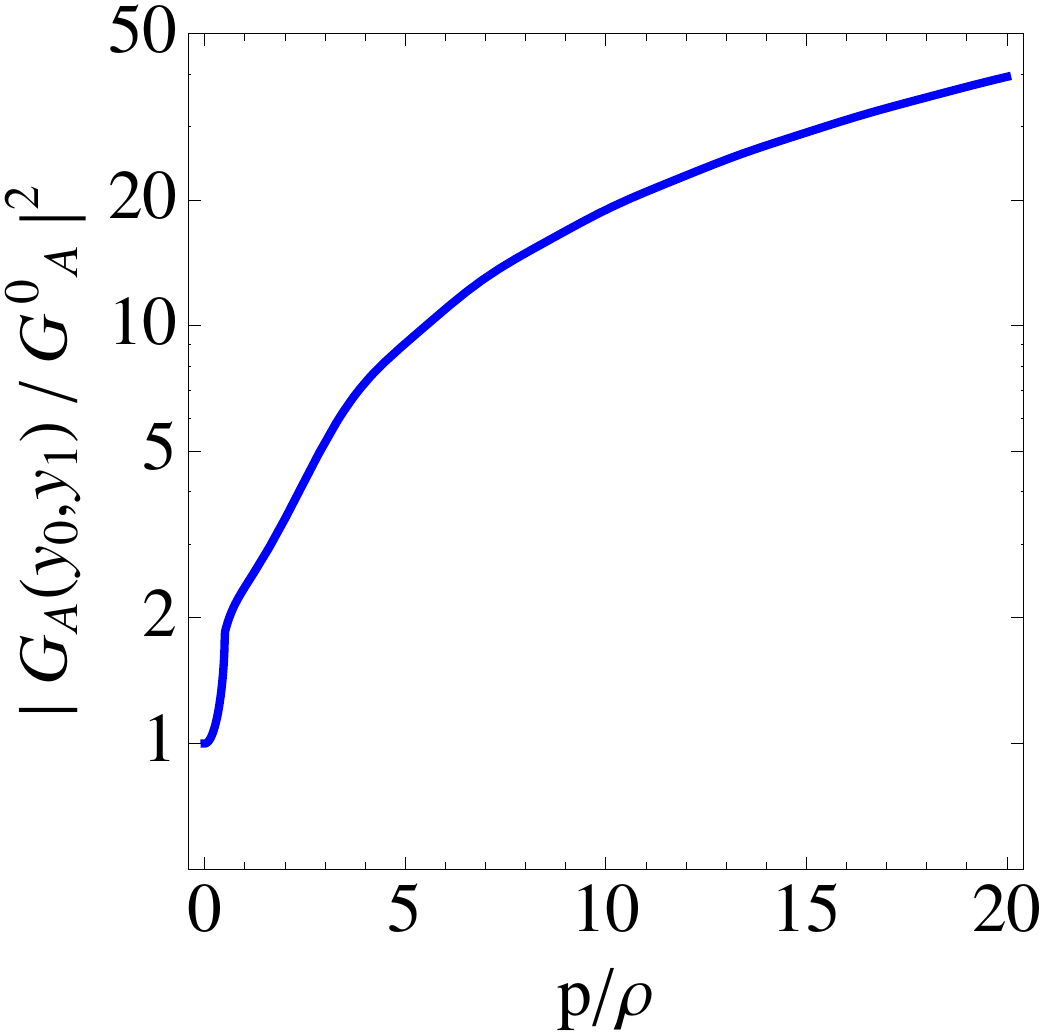} \hspace{0.5cm}
\includegraphics[width=4.9cm]{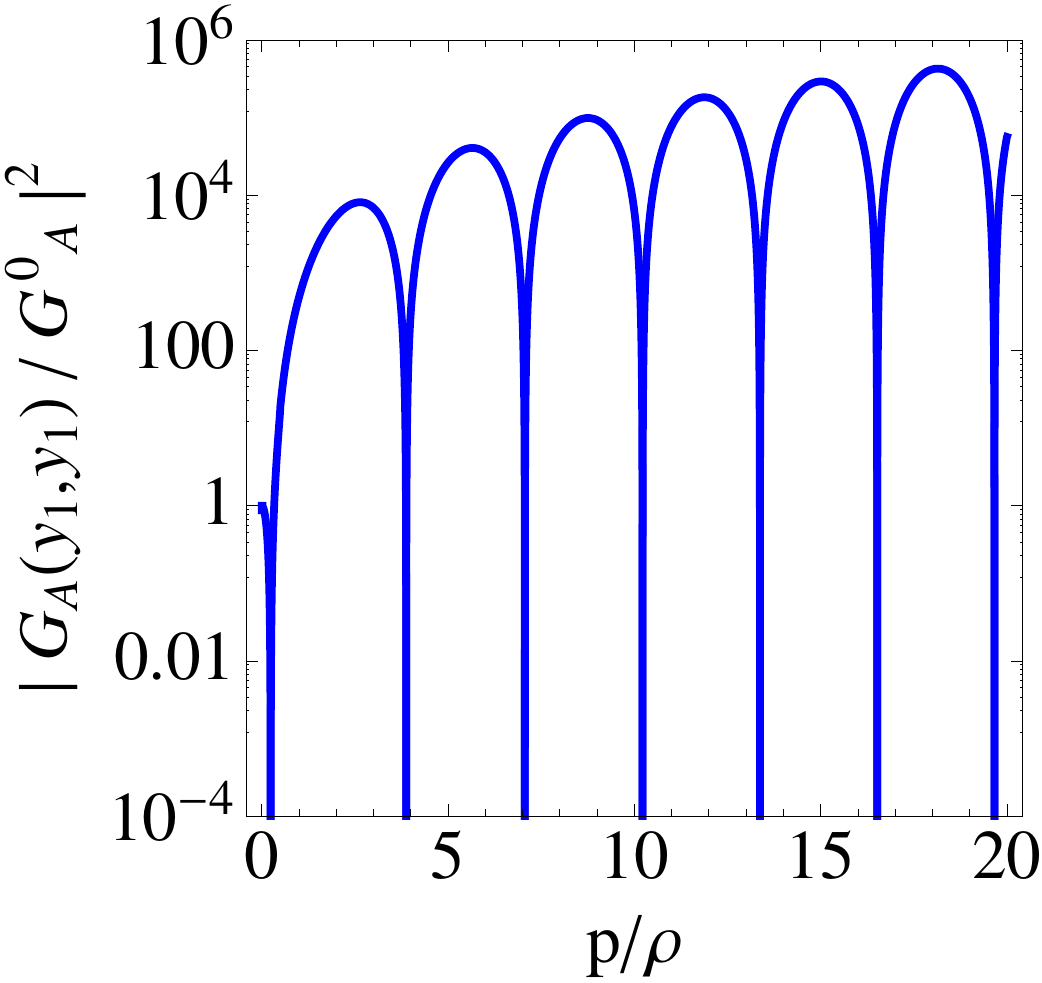}
\vspace{-0.2cm}
\caption{\it Plots of $|G_A(y_0,y_0;p)/G_A^0|^2$ (left panel), $|\mathcal G_A(y_0,y_1;p)/G_A^0|^2$ (middle panel), and $|G_A(y_1,y_1;p)/G_A^0|^2$ (right panel) as functions of $p/\rho$.  These quantities can be interpreted as $\sigma(q\bar q\to g^*\to f_{X}\bar f_{X}) / \sigma_{\SM}( q\bar q \to g^{(0)}\to  f_{X}\bar f_{X} )$, where $f_{X}$ is a light quark living on the UV brane as e.g. $b_{L,R}$  (left panel) or a heavy quark living on the IR brane as e.g. $t_R$ (middle panel), and  $\sigma(b_R\bar b_R \to g^*\to t_{R} \bar t_{R}) / \sigma_{\SM}( b_R \bar b_R \to g^{(0)}\to  t_{R} \bar t_{R} )$ (right panel), cf. Sec.~\ref{sec:phenomenology}. We have used $A_1 = 35$ in all panels and assume time-like momenta $p^2>0$.}
\label{fig:GA}
\end{figure*} 

The Green's functions that we will study generalize the particle propagators with isolated poles
\begin{equation}
\frac{1}{p^2 - m^2 + i 0^+} = \mathcal{P} \frac{1}{p^2 - m^2} - i\pi \delta(p^2 - m^2) \,,
\end{equation}
to Green's functions with an isolated pole (zero mode) and a continuum of states (instead of a discrete sum of KK modes) with a mass gap $m_g$. This is the behavior of gapped unparticles. In the present model, the gap $m_g \sim$ TeV is linked to the solution of the hierarchy problem.

\subsection{General Green's functions}
\label{subsec:Green_gauge_bosons}

The Lagrangian for massless gauge bosons writes as
\begin{equation}
\mathcal L = \int_0^{y_s} dy\left[ -\frac{1}{4} \tr F_{\mu\nu}F^{\mu\nu}-\frac{1}{2}e^{-2A} \, \tr A'_\mu A'_\mu  \right]\,,
\end{equation}
where the gauge fluctuations can be written as $A_\mu(p,y) = f_A(y) A_\mu(p) / \sqrt{y_s}$. The EoM of the fluctuations can be expressed in a Schr\"odinger like form  after rescaling the field, leading to the effective potential
\begin{equation}
V_A(z) =  \frac{3}{4z^2}  \Theta(z_1 - z) +  m_g^2 \Theta(z - z_1) \,, \label{eq:VAModelII}
\end{equation}
with $m_g = \rho/2$. The Green's functions for gauge bosons propagating in the bulk from $y$ to $y^\prime$ obey the~EoM
\begin{equation}
\hspace{-0.5cm} p^2 G_A(y,y^\prime;p) + \partial_y \left( e^{-2A(y)} \partial_y G_A(y,y^\prime;p) \right) = \delta(y-y^\prime) \,,  \label{eq:Ghy}
\end{equation}
that can be solved analytically by imposing Neumann boundary conditions in the UV brane and regularity in the IR singularity, cf. Ref.~\cite{Megias:2021arn} for details. All Green's functions include a zero mode contribution
\begin{equation}
G_A^0 =  \frac{1}{y_s p^2} = \lim_{p \to 0} G_A(y,y^\prime;p)  \,.
\end{equation}
We plot in Fig.~\ref{fig:GA} the normalized brane-to-brane Green's functions $|G_A(y_\alpha,y_\beta;p)/G_A^0|^2$. The phenomenological consequences of this behavior will be studied in Sec.~\ref{sec:phenomenology}.

\subsection{Resonances}
\label{subsec:resonances}

It is well known in Quantum Field Theory (QFT) that resonances with mass $M_n$ and decay width $\Gamma_n$ are associated to the presence of poles in the second Riemann sheet of the complex $s \equiv p^2$ plane at
\begin{equation}
s_n = M_n^2-i M_n \Gamma = M_n^2(1-i r_n),\quad (r_n \equiv \Gamma_n/M_n)  \,. 
\end{equation}
While, in the case of unparticles, it is not clear whether these poles are associated to decays of resonances, it is worth exploring them in the gapped continuum model, as the breaking of conformal invariance by the branes makes a difference with respect to unparticles.

\begin{figure}[htp] 
\centering
\includegraphics[width=4.8cm]{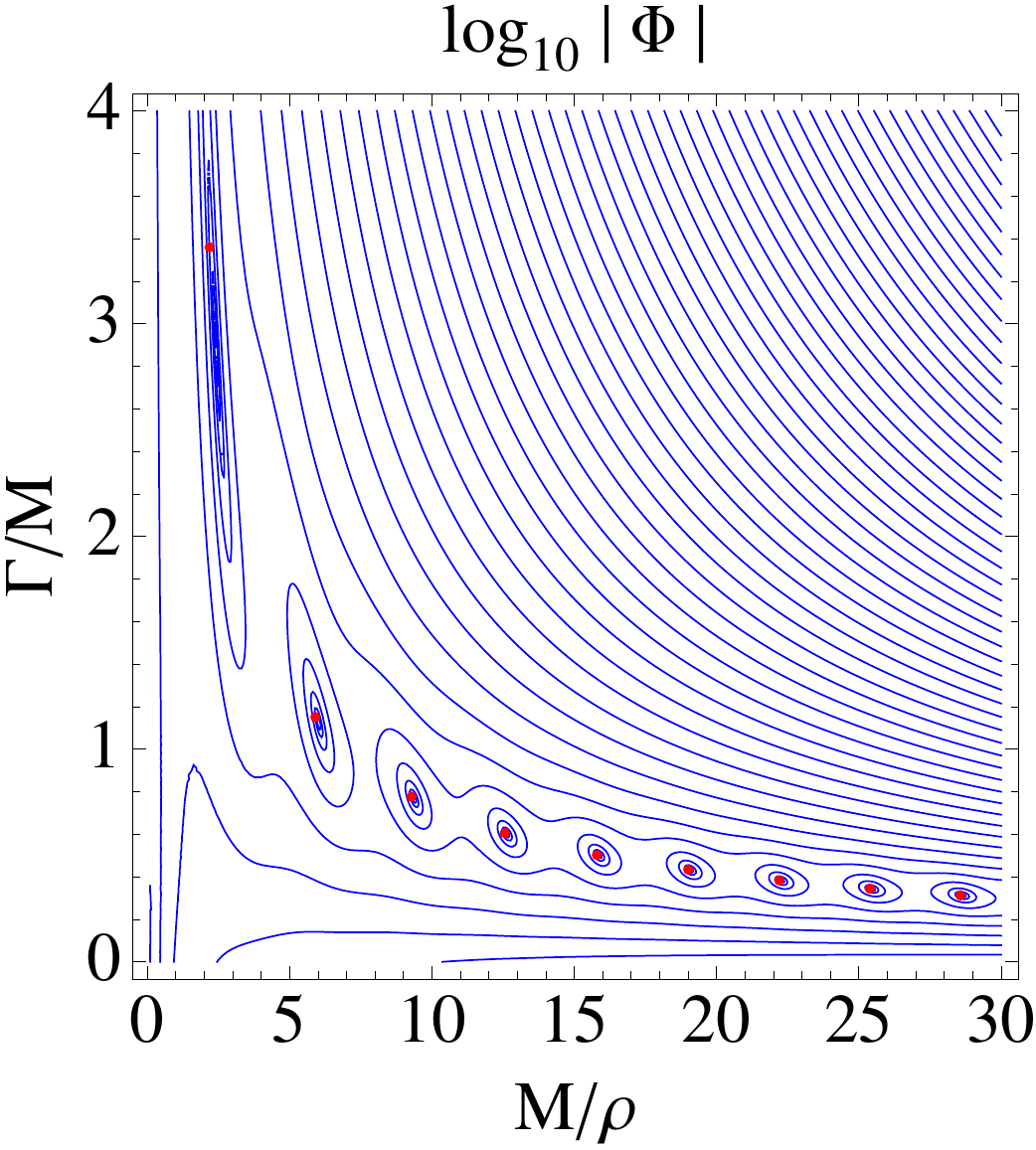} 
\vspace{-0.2cm}
\caption{\it Contour plot of $\log_{10} |\Phi(p)|$ in the plane $(M/\rho, \Gamma/M)$. The contour lines form small circles around the zeros of $\Phi(p)$. The (red) dots stand for the positions of the zeros of $\Phi(p)$ as given by Eq.~(\ref{eq:mn_analytical}).}
\label{fig:resonances}
\end{figure}

The origin of the imaginary part of the Green's functions is the threshold function $\delta_A(s) = (1-s/m_g^2)^{1/2}$ which has two Riemann sheets. The change from the first (I) to the second (II) Riemann sheet is equivalent to $\delta_A^{\rm II}(s) = -\delta_A^{\rm I}(s)$. On the other hand, all the Green's functions are $G_A(y,y^\prime;p) \propto 1/\Phi(p)$, where $\Phi(p)$ is a universal function, so that the zeros of $\Phi(p)$ correspond to the poles of $G_A$. Following this idea, we display in Fig.~\ref{fig:resonances} a contour plot of $\log_{10} |\Phi(p)|$ in the second Riemann sheet. There appears an intriguing structure of zeros. The values of $M_n/\rho$ follow a pattern similar to the KK modes in the RS model, but contrary to that model, the resonances have a finite width. 

By studying the behavior of $\Phi(p)$ at large momentum, it is possible to derive an approximate analytical formula for the position of the poles. Using this procedure, the zeros of this function turn out to be~\cite{Megias:2021arn}
\begin{equation}
\frac{s}{\rho^2} \simeq -\mathcal W_n\left[ \pm \frac{1}{4}(1+i) \right]^2  \,, \;  n = -1,-2, \cdots \,, \label{eq:mn_analytical}
\end{equation}
where $\mathcal W_n(z)$ is the $n$-th branch of the Lambert function.

\vspace{-0.5cm} 

\section{Spectral functions}
\label{sec:spectral_function}

\begin{figure*}[t]
\centering
\includegraphics[width=4.8cm]{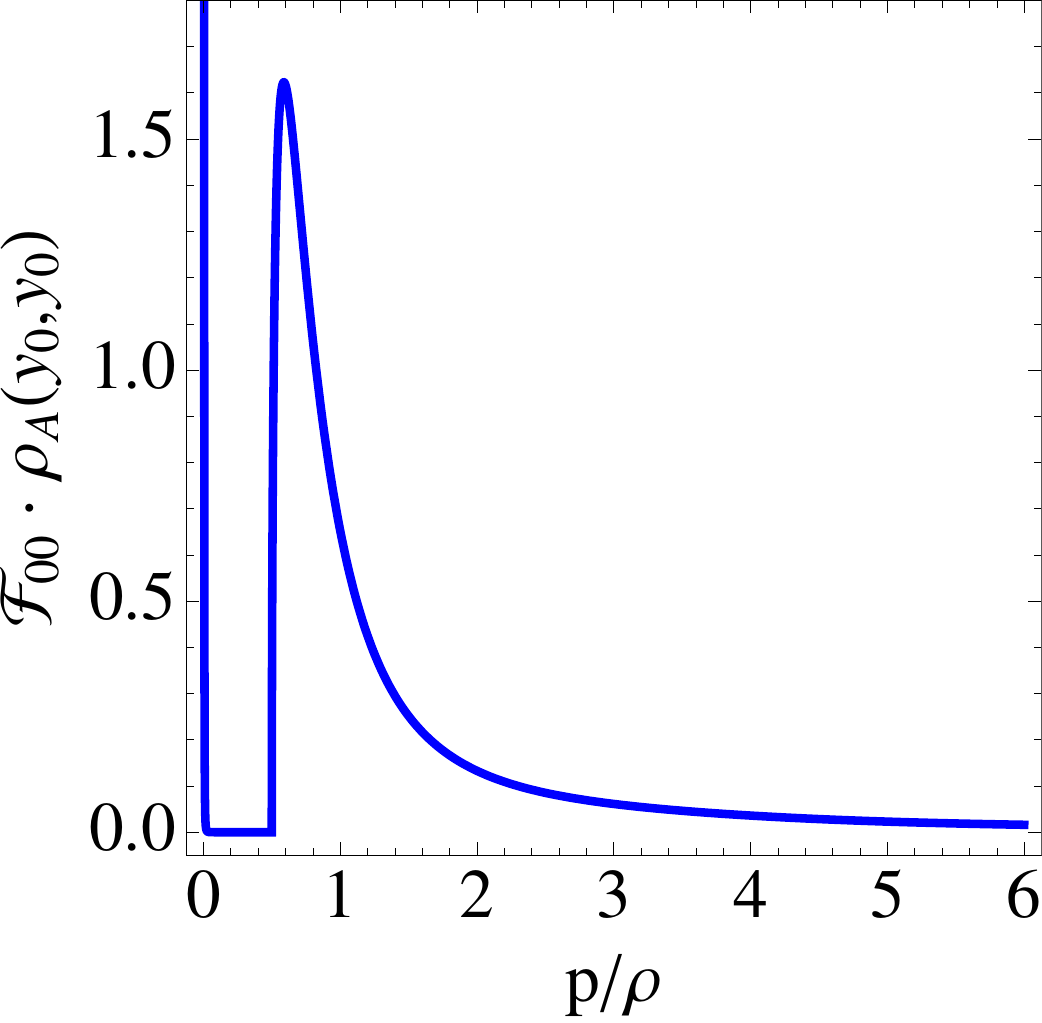} \hspace{0.5cm}
\includegraphics[width=5.1cm]{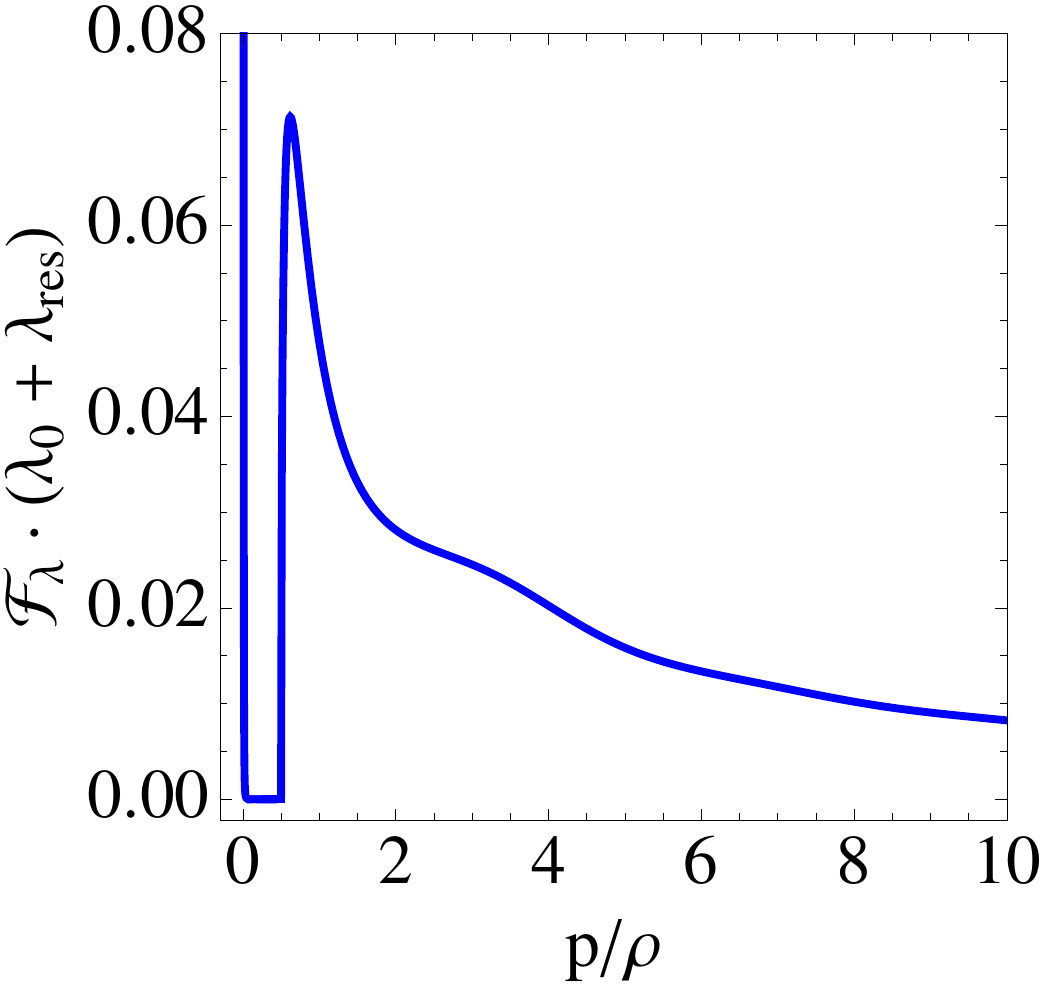} \hspace{0.5cm}
\includegraphics[width=4.9cm]{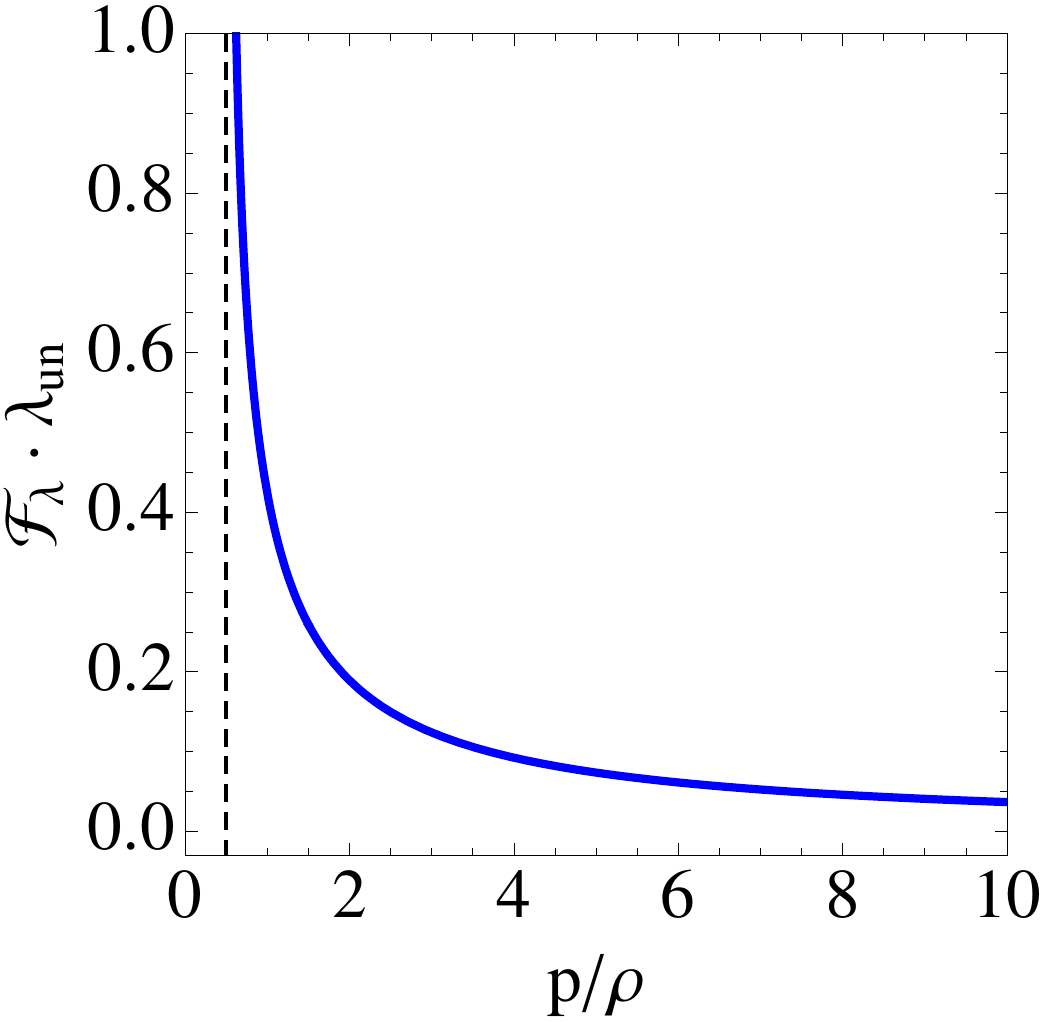} 
\vspace{-0.2cm}
\caption{\it Left panel: Spectral function $\mathcal F_{00} \cdot \rho_A(y_0,y_0;p)$ as a function of $p/\rho$.  Middle and right panels: Regular $\mathcal F_\lambda \cdot \left( \lambda_0 + \lambda_{\textrm{res}} \right)$ (middle panel) and unparticle $\mathcal F_\lambda \cdot \lambda_{\textrm{un}}$ (right panel) contributions to the eigenvalue $\lambda$ in the gapped continuum model, cf. Eqs.~(\ref{eq:lambda_0})-(\ref{eq:lambda_un}), with $\mathcal F_\lambda = \rho^2$. The dashed vertical line in the right panel corresponds to $p = m_g$. We have used $\epsilon = 0.1/k$, $A_1 = 35$ and assume time-like momenta $p^2>0$.}
\label{fig:spectral}
\end{figure*}

The spectral functions are defined as
\begin{equation}
\rho_A(y,y^\prime;s)= - \frac{1}{\pi} \textrm{Im } G_A(y,y^\prime;s + i0^+) \,.
\end{equation}
They are related to a discontinuity of $G_A$ at the branch cut $s \in [m_g^2 , +\infty)$ if one approaches it using the same Riemann sheet. An explicit evaluation leads to
\begin{equation}
\rho_A(y,y^\prime;s) =  \frac{1}{y_s} \, \delta(s) +  \eta_A(y,y^\prime;s) \Theta(s - m_g^2) \,,
\end{equation}
containing a zero mode, and a continuum contribution above the gap.  We show $\mathcal F_{00} \cdot \rho_A(y_0,y_0;p)$ in Fig.~\ref{fig:spectral} (left),  where $\mathcal F_{00} \equiv \frac{\rho^2}{k} (ky_s)^{2}$ makes it approximately invariant under changes of~$A_1$. Let us point out  that $\rho_A(y,y^\prime)$ with $y \ne y^\prime$ is not positive definite, a fact that challenges the physical interpretation of the spectral functions in 4D QFT. We will address below this apparent contradiction.

\subsection{Positivity of the spectral function}
\label{subsec:positivity}

The spectral function $\rho_A(y,y^\prime;s)$ can be understood as a matrix element of a spectral operator
\begin{equation}
\hspace{-0.5cm} \hat \rho_A = - \frac{1}{\pi} \Imaginary \; \hat G_A \,, \;\, \textrm{where} \; \Imaginary \;\, \hat G_A = \frac{1}{2i} \left( \hat G_A - \hat G_A^\dagger \right)  \,,
\end{equation}
i.e. $(\hat \rho_A)^{y}_{y^\prime} \equiv \rho_A(y,y^\prime;s) = \langle y |\, \hat \rho_A | y^\prime \rangle$, where we have~used the Dirac notation. Then, it can be proved that $\hat\rho_A$ is positive semidefinite. To see that, let us point out that the infinite dimensional matrix $\hat\rho_A$ has a factorizable form $(\hat \rho_A)^{y}_{y^\prime} = \rho_y \rho_{y^\prime}$, where $\rho_y = \sqrt{(\hat\rho_A)^y_y}$, so that all its eigenvalues are zero except one, which is given by its trace,
\begin{equation}
\hspace{-0.5cm} \lambda(s) = \tr\, \hat\rho_A = \int_0^{y_s} dy \, \rho_{A}(y,y;s) = \int_0^{y_s} dy \, \rho_y^2  \ge 0  \,. \label{eq:lambda_s}
\end{equation}
We will compute below this eigenvalue in the RS model, for illustration, and in the gapped continuum model.

\subsection{Spectral function in the RS model}
\label{subsec:spectral_RS}

The spectral function in the RS model $(A(y) = ky, \; 0 \le y \le y_1)$ can be written as
\begin{equation}
\hspace{-0.4cm} \rho_{A,\RS}(y,y^\prime;s) = \frac{1}{y_1} \sum_n f_A^{(n)}(y) f_A^{(n)}(y^\prime)^\ast  \; \delta(s - m_n^2)   \,, \label{eq:rho_RS}
\end{equation}
where $f_A^{(n)}(y)$ is the $n$-th KK mode wave function. The spectrum is approximately given by the zero mode and the zeros of the Bessel function $J_0(m/\rho)$, i.e. $m_n/\rho \simeq 0,\,  2.4, \, 5.6, \, 8.7, \dots$. Using $\int_0^{y_1} \! dy |f_A^{(n)}(y)|^2 \!=\! y_1$, one finds
\begin{equation}
\lambda_{\RS}(s) =  \sum_{n} \delta(s - m_n^2)  \ge 0 \,,
\end{equation}
so that $\lambda_{\RS}(s)$ can be interpreted as the {\it density of states} in momentum $s$. On the other hand, the integral
\begin{equation}
\hspace{-0.7cm} \int_0^\infty \!\! ds \, \lambda_{\RS}(s) = \int_0^\infty \!\! ds \sum_n \delta(s - m_n^2) = N_{\textrm{states}} \to \infty 
\end{equation}
is the {\it number of states}, a quantity that turns out to be divergent as there are infinite states in the spectrum.

\subsection{Spectral function in the gapped continuum model}
\label{subsec:spectral_gapped_continuum}

This model leads to $\lambda(s)  = \lambda_0(s) + \lambda_{\textrm{res}}(s) + \lambda_{\textrm{un}}(s)$ with
\begin{eqnarray}
\hspace{-0.5cm} &&\lambda_0(s) = \delta(s) \,, \quad \lambda_{\textrm{res}}(s) = \xi_A(s) \Theta(s - m_g^2) \,, \label{eq:lambda_0} \\
\hspace{-0.5cm} &&\lambda_{\textrm{un}}(s) =  -\frac{\log(k\epsilon)}{2\pi \rho (s - m_g^2)^{1/2}} \Theta(s - m_g^2) \,, \label{eq:lambda_un}
\end{eqnarray}
where $\xi_A(s)$ is a continuous positive function vanishing at $s = m_g^2$, and having a maximum close to $m_g^2$, while $\epsilon$ is a cutoff in the IR singularity, so that the integral in Eq.~(\ref{eq:lambda_s}) is up to $y_s - c \cdot \epsilon$ with $c \simeq 1.62$.  These contributions correspond, respectively, to the zero mode, a term affected by the resonances of Sec.~\ref{subsec:resonances}, and a contribution characterized by the absence of resonances. The latter, which is divergent in the limit $s \to m_g^{2 \, +}$, corresponds to a contribution from unparticles with a dimension $d_U = 3/2$ and a mass gap, see e.g. Ref.~\cite{Delgado:2008gj}. In Fig.~\ref{fig:spectral} the regular $\lambda_0 + \lambda_{\textrm{res}}$ (middle panel) and the unparticle $\lambda_{\textrm{un}}$ (right panel) contributions are displayed.

\section{Phenomenological aspects}
\label{sec:phenomenology}

The main experimental signature for new physics is its production, and subsequent decay, in colliders. When the new physics consists in a continuum of states, its presence should be associated with an excess with respect to the SM prediction in the measured cross section. We will address below how the existence of a continuum can affect the cross sections in $pp$ collisions.

\subsection{UV-to-UV/IR brane Green's function}
\label{subsec:UV_IR_Green}

In Drell-Yan processes the continuum of KK gluons is produced by pairs of light fermions  $(q\bar q)$ localized on the UV brane. The continuum will decay into a pair of light/heavy fermions $f_{\UV} \bar f_{\UV}/f_{\IR} \bar f_{\IR}$ localized on the UV/IR brane. The contribution of the continuum gauge bosons $g^\ast$ produces an excess in the cross section
\begin{equation}
\frac{\sigma(q\bar q\to g^*\to f_{X}\bar f_{X}) }{ \sigma_{\SM}( q\bar q \to g^{(0)}\to  f_{X}\bar f_{X} ) } = \left| \frac{G_A(y_0,y_X;\hat s) }{ G_A^0 } \right|^2  \,,  \label{eq:sigma_q}
\end{equation}
where $f_X = f_{\UV}/f_{\IR}$ and $y_X = y_0/y_1$, $p = \sqrt{\hat s}$ is the partonic center of mass energy, $g^{(0)}$ is the SM gluon, and $q$ is a proton valence light quark. It was displayed in the left and middle panels of Fig.~\ref{fig:GA} the cases $f_{\UV} = b_{L,R}$ and $f_{\IR} = t_R$, respectively. As fermions localized on the IR brane are strongly coupled to the KK modes, the cross section yields an enhancement $\mathcal O(20)$ for $p/\rho \simeq 10$ in the case $f_{\IR} = t_{R}$, while it is $\mathcal O(1)$ for $f_{\UV} = b_{L,R}$.

\subsection{IR-to-IR brane Green's function}
\label{subsec:IR_IR_Green}

The Green's function $G_A(y_1,y_1;\hat s)$ is relevant in processes where both the initial and final fermions are localized on the IR brane. This is the case in models explaining the $R_{D^{(\ast)}}$ anomalies~\cite{Megias:2017ove,Carena:2018cow}. The ratio
\begin{equation}
\frac{\sigma( b_R\bar b_R \to g^*\to t_R \bar t_R) }{ \sigma_{\SM}( b_R\bar b_R \to g^{(0)}\to t_R\bar t_R) } = \left| \frac{G_A(y_1,y_1;\hat s)}{G_A^0} \right |^2 , \label{eq:sigma_bR}
\end{equation}
is significant due to the large coupling of $b_R$ to the KK gluon modes. Fig.~\ref{fig:GA} (right) shows that the enhancement of the production can easily be $\mathcal O(10^4-10^6)$. Finally, let us mention that all these processes, Eqs.~(\ref{eq:sigma_q})-(\ref{eq:sigma_bR}),  are dominated in the limit $p \to 0$ by the gluon zero mode.

\section{Conclusions}
\label{sec:conclusions}

We have studied a 5D warped model that solves the hierarchy problem. The metric of the model has an AdS$_5$ behavior between the UV and the IR brane, and that of a LDM near the IR singularity. Within this model, we find that the KK spectrum of SM massless gauge bosons (photon and gluon) has an isolated massless pole corresponding to the 4D gauge boson, and a continuum with a mass gap $m_g = \rho/2$ where $\rho \sim \TeV$.

When considering the Green's functions in the complex $s$ plane, we find the existence of poles in the second Riemann sheet, indicating the presence of broad resonances. Let us stress that these resonance effects already appear at tree level. We have computed also the spectral functions and studied their positivity. Finally, we have analyzed how the existence of continuum spectra can modify the present searches of new physics. 

This work can be extended to the computation of the Green's functions of other fields (fermions, \dots), as well as to the study of the couplings of the continuum KK modes with the SM fields. These and other applications, that can be inspired on unparticle phenomenology, will be addressed in a forthcoming publication~\cite{Megias:prep}.

\section*{Acknowledgements} 
We would like to thank A.~Carmona, M.~P\'erez-Victoria and
L.L.~Salcedo for fruitful discussions. The work of EM is supported by
the Spanish MINECO under Grants FIS2017-85053-C2-1-P and
PID2020-114767GB-I00, by the FEDER/Junta de
Andaluc\'{\i}a-Consejer\'{\i}a de Econom\'{\i}a y Conocimiento
2014-2020 Operational Programme under Grant A-FQM-178-UGR18, by Junta
de Andaluc\'{\i}a under Grant FQM-225, and by the Consejer\'{\i}a de
Conocimiento, Investigaci\'on y Universidad of the Junta de
Andaluc\'{\i}a and European Regional Development Fund (ERDF) under
Grant SOMM17/6105/UGR. The research of EM is also supported by the
Ram\'on y Cajal Program of the Spanish MINECO under Grant
RYC-2016-20678. The work of MQ is partly supported by the Spanish
MINECO under Grant FPA2017-88915-P, by the Catalan Government under
Grant 2017SGR1069, and by Severo Ochoa Excellence Program of MINECO
under Grant SEV-2016-0588. IFAE is partially funded by the CERCA
program of the Generalitat de Catalunya.

\end{document}